Letter to the Editor

# Eruption of magnetic flux ropes during flux emergence

V. Archontis[1] and T. Török[2]

[1] School of Mathematics and Statistics, University of St. Andrews, Noth Haugh, St. Andrews, Fife, KY16 9SS, UK
  e-mail: vasilis@mcs.st-and.ac.uk
[2] LESIA, Observatoire de Paris, CNRS, UPMC, Université Paris Diderot, 5 place Jules Janssen, 92190 Meudon, France
  e-mail: tibor.torok@obspm.fr

**ABSTRACT**

*Aims.* We investigate the formation of flux ropes in a flux emergence region and their rise into the outer atmosphere of the Sun.
*Methods.* We perform 3D numerical experiments solving the time-dependent and resistive MHD equations.
*Results.* A sub-photospheric twisted flux tube rises from the solar interior and expands into the corona. A flux rope is formed within the expanding field, due to shearing and reconnection of field lines at low atmospheric heights. If the tube emerges into a non-magnetized atmosphere, the flux rope rises, but remains confined inside the expanding magnetized volume. On the contrary, if the expanding tube is allowed to reconnect with a preexisting coronal field, the flux rope experiences a full eruption with a rise profile which is in qualitative agreement with erupting filaments and Coronal Mass Ejections.

**Key words.** Magnetohydrodynamics (MHD) – Methods: numerical – Sun: activity – Sun: corona – Sun: magnetic fields

## 1. Introduction

It is now widely accepted that filament eruptions and coronal mass ejections (CMEs), as well as associated flares, are not distinct phenomena, but different observational manifestations of a sudden and violent destabilization of the coronal magnetic field. This destabilization is believed to occur when the field has been stressed by photospheric motions or newly emerging flux to a point where it cannot longer maintain equilibrium and erupts (Forbes 2000). The eruption can be confined in the low corona, i.e. not developing into a CME, if the ambient magnetic field overlaying the erupting core flux does not drop sufficiently fast with height (Török & Kliem 2005; Liu 2008). Typically, CMEs exhibit three distinct phases (Zhang et al. 2001): an initiation phase, often observed as the slow rise of a filament (e.g. Sterling & Moore 2005; Schrijver et al. 2008), a rapid acceleration phase, and a propagation phase at approximately constant velocity. The acceleration phase is in most cases characterized by an exponential-like rise of the erupting structure (Vršnak 2001). There are two closely coupled mechanisms which are considered to drive the eruption in this phase: an ideal MHD flux rope catastrophe or instability (e.g. Forbes & Isenberg 1991; Kliem & Török 2006) and "runaway reconnection" due to the feedback between the CME expansion and the associated reconnection below it (e.g. Vršnak et al. 2004, and references therein).

CMEs are known to originate predominantly in active regions and are often associated with flux emergence. They can occur during the whole lifetime of an active region, from its emergence to its decay (e.g. Green et al. 2002). In many cases, the emerging flux merely acts as a trigger for the eruption of an already existing filament (Feynman & Martin 1995; Williams et al. 2005). On the other hand, the occurrence of CMEs in active regions in their early phase of formation (e.g. Démoulin et al. 2002; Nindos et al. 2003) suggest that emerging flux carries the potential to produce eruptions by itself.

In numerical experiments, magnetic flux rope models have been extensively used to study solar eruptions. In this framework, Manchester et al. (2004) simulated the emergence of a twisted flux tube into a non-magnetized, stratified atmosphere and they reported the formation of a distinct magnetic flux rope within the emerging tube when the length of the buoyant part of the tube is small. The shearing of the magnetic field along the polarity inversion line of the active region played a key role on the formation of the rope. The rope rised into the atmosphere with a moderate speed and eventually left the simulation box through the open top boundary. The authors did not associate this rising motion with a filament eruption or CME due to the lack of a quiescent state preceding the rise of the flux rope and the lack of sufficient "axial flux" to drive a full eruption. Also, Gibson & Fan (2006) studied the eruption of a rope when enough twist has emerged to induce a loss of equilibrium. In this case, a flux rope can break in two, with the upper part leaving the corona.

The simulations presented in this Letter suggest that flux rope formation in an emerging flux region is a generic phenomenon, independent of parameters such as the twist and the length of the buoyant part of the tube. The rope erupts into the outer atmosphere when the primary emerging field reconnects with pre-existing coronal field; otherwise it remains confined by the ambient field of the initial emerging system. The eruption of the rope is in agreement with the observed character of height-time profiles in filament eruptions and CMEs.

## 2. Model

The setup of our experiments is very similar to Archontis et al. (2005): we integrate the resistive MHD equations in a cartesian box of dimensionless size $[-70, 70] \times [-80, 80] \times [-10, 110]$ in the $x, y, z$ directions respectively. We impose a non-magnetic background stratification including a subphotospheric domain ($-10 < z < 20$), photosphere ($20 < z < 30$), transition region ($30 < z < 40$), and corona ($40 < z < 110$). In the experiment described in section 3.2, we add a horizontal magnetic field in the transition region and corona. We use periodic boundary conditions in horizontal directions and closed boundaries with a wave damping layer in vertical directions. All variables are made dimensionless by choosing photospheric values for density, $\rho_{ph} = 3 \times 10^{-7}\,\mathrm{g\,cm^{-3}}$, pressure, $p_{ph} = 1.4 \times 10^5\,\mathrm{ergs\,cm^{-3}}$, and pressure scale height, $H_{ph} = 170\,\mathrm{km}$, and by derived units (e.g., magnetic field strength $B_{ph} = 1300\,\mathrm{G}$, velocity $V_{ph} = 6.8\,\mathrm{km\,s^{-1}}$ and time $t_{ph} = 25\,\mathrm{s}$). A horizontal, uniformly twisted flux tube of radius $R = 2.5$ is placed at $z = 12$ below the photosphere, oriented along the $y$ axis. It is defined by $B_y = B_0 \exp(-r^2/R^2)$ and $B_\phi = \alpha\, r\, B_y$, where $r$ is the radial distance from the tube axis and $\alpha$ is the twist per unit length. The tube is made buoyant by applying a density perturbation $\Delta\rho = [p_t(r)/p(z)]\rho(z)\exp(-y^2/\lambda^2)$, where $p_t$ is the pressure within the flux tube. We use $\lambda = 10$ throughout this paper.

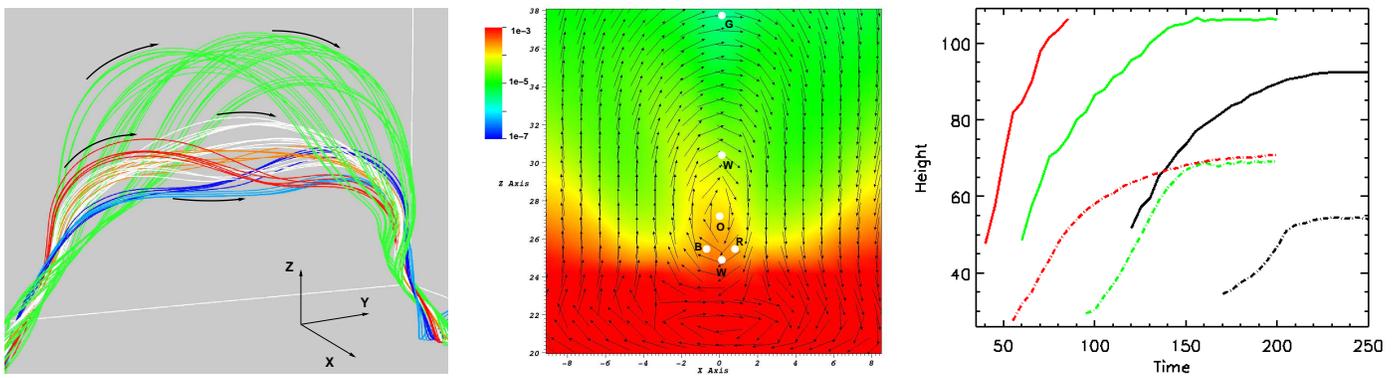

**Fig. 1.** *Left:* Magnetic field lines in experiment E1 at $t = 55$. The arrows show the direction of the magnetic field vector. *Centre:* Colormap showing the density and, superimposed, the projection of the magnetic field onto the $y = 0$ plane. The letters correspond to the color of the field lines that pass through the specific points (e.g R for red and O for orange). *Right:* Height-time profiles of the flux rope centre *(dot-dashed)* and the top of the expanding ambient field *(solid)* in experiments E1 *(red)*, E2 *(green)* and E3 *(black)*.

## 3. Results

### 3.1. Emergence into a field-free corona: confined eruption

We consider three experiments, with $B_0 = 5$ and $\alpha = 0.4$ (E1), $B_0 = 3$ and $\alpha = 0.4$ (E2), and $B_0 = 3$ and $\alpha = 0.1$ (E3), respectively. In all experiments, the tube rises from the solar interior to the photosphere, where it slows down. The further rise into the atmosphere occurs through the development of a magnetic buoyancy instability (Archontis et al. 2004). As the tube rises, it substantially expands due to the marked decrease in background gas pressure with height. The outermost *external* magnetic field lines are strongly azimuthal if the initial flux tube twist is strong. However, *internal* field lines, which are closer to the tube axis, are oriented almost parallel to the axial field.

When the rising field reaches the photosphere, a bipolar region is formed. As the two polarities move in opposite directions, the field becomes sheared along the neutral line. The middle part of the emerging field starts to rise at a faster rate as the lateral field, which is more dense. Thus, a local decrease of magnetic (and total) pressure along the transverse direction at low heights develops. As a result, the flanking magnetized plasma is forced by the resulting pressure gradient to move inward. This inflow, together with the horizontal shearing, brings the internal field lines together and they reconnect to form a weakly twisted *flux rope* above the axis of the emerging tube, between the upper atmosphere and the transition region. A *sigmoidal* current layer forms between the flux rope and the tube axis.

The left panel in Figure 1 shows the field line topology at an early stage of rope formation (case E1). The blue and red field lines are sheared on opposite sides of the polarity inversion line. Above the photosphere, they have a convex arch-like shape at one end and a slight concave upward (dip-like) shape at the other end. The central segments of these two sets of field lines are oriented in opposite directions. The red ones are directed downward and the blue ones upward. They reconnect, forming the orange lines that pass through the central region of the flux rope, and the cyan field lines, which overlay the top of the original emerging axial field. White field lines have been traced above and below the flux rope centre. They wrap around the rope center in opposite directions. The lower ones make a full rotation above the photosphere, possessing dips at their middle section. The upper ones are convex at the top, exhibiting a loop-shape. Green field lines represent the top part of the expanding flux system, which rises unimpeded into the non-magnetized corona and, thus, forms an *ambient* magnetic field for the flux rope. The central panel shows the density distribution in the $y = 0$ vertical plane. Arrows show the projection of the magnetic field vector on this plane, outlining the flux rope. An essential feature is that the field lines below the rope axis have dips where dense plasma is trapped. After its formation, the flux rope starts to rise and the dense plasma is transported into the corona.

The right panel in Fig. 1 shows height-time profiles of the flux rope and the ambient field in our three experiments. The emerging tube in E1 reaches photospheric heights more quickly, since the buoyancy force in the central part of the tube is proportional to $B_0^2$. Thus, the tube has larger momentum when reaching the photosphere and start to rise into the upper atmosphere earlier than in E2 and E3. The rate of emergence (i.e. the flux transported into the atmosphere per time unit) is also higher in E1. Due to the smaller $B_0$ and $\alpha$, the magnetic field strength and gradient at the tube apex is smaller in E3 than in E1 and E2, when it arrives at the photosphere. Thus, the plasma $\beta$ is larger, the buoyancy instability is not as easily triggered and the emergence of the field and subsequent formation of the flux rope is significantly delayed compared to E1 and E2. For a detailed parametric study identifying the effects of different twist and field strength in emerging flux tubes see Murray et al. (2006).

The rise of the rope is measured by locating its axis at different times. To that end, we locate the center (or 0-point) of the island-like cross section of the rope at the x-z vertical midplane.

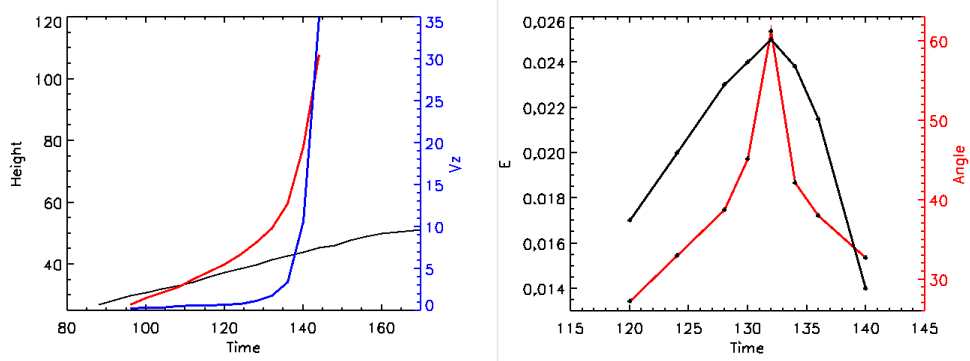

**Fig. 2.** *Left:* Evolution of the height of the flux rope axis in two experiments: without preexisting field (black line) and with preexisting field (red line). The velocity of the rope in the latter case (blue line) is drawn against the right y-axis. *Right:* Electric field (black) and contact angle (red) in the experiment with preexisting field.

We do not find a change in the field line orientation, during the rising motion, nor a significant rotation of the rope about the z-axis and, thus, the 0-point outlines the axis of the rope at a very good approximation. During their evolution, the flux ropes accelerate to reach a modest maximum velocity of $V_z \approx 2.8$. The maximum speed is reached $\approx 30$ time units after the rope formation. The dominant force that drives the rise of the rope is the magnetic pressure force. Also, the magnetic pressure inside the rope is larger than the background gas pressure. The latter leads to an expansion of the rope as it rises into the atmosphere, which in turn reduces the magnetic field strength within the rope. Eventually, the flux ropes slow down and their rise saturates at coronal heights. The saturation sets in all cases roughly at the same time when the expansion of the ambient field stops. In E1 and E2, the expansion stops since the ambient field reaches the top of the simulation box. In a larger domain, the ambient field and the rope reach larger heights. In E3, however, the expansion stops already when the apex of the ambient field has reached $z \approx 90$. This is because the magnetic pressure of the expanding field in E3 is small enough to be balanced by the external gas pressure at this stage of the evolution. At the onset of deceleration, the flux ropes have adopted an arch-like shape and the downward tension force of the field lines increases. Eventually, the tension nearly balances the gradient of the magnetic pressure and the rope approaches an equilibrium. At the time the ambient field expansion stops, the magnetic field strength drops only slightly with height above the flux rope. As a result, the magnetic tension of the overlying field stops the rise of the rope (see also Török & Kliem 2005).

### 3.2. Emergence into a magnetized corona: full eruption

The confinement of the flux rope may fail if the ambient field drops faster with height above the flux rope, or is removed (e.g., via reconnection). To see this, we perform another experiment with $B_0 = 5$ and $\alpha = 0.4$, but now we add a weak, horizontal, space-filling magnetic field of strength $5 \cdot 10^{-3}$ to the initial upper atmosphere. The field is uniform in the corona and transition region and drops to zero between $z = 10$ and $z = 20$. We choose its orientation nearly antiparallel to the top of the expanding ambient field, hence favourable for reconnection. In these experiments the size of the numerical domain is $(-50, 50)$, $(-50, 50)$, $(-10, 110)$ in the $x, y, z$ directions respectively.

Figure 2 shows height-time profiles of the flux rope in two experiments, with and without preexisting field (red and black line respectively). In the case without ambient field the rope rises linearly with time. The slower rate of the rising motion of the rope (compared to E1) is due to the smaller size of the numerical domain, which restricts the shearing and expansion of the field and, thus, bounds the general evolution of the system. With preexisting field, the rope rises much higher into the atmosphere, showing a two-phase evolution: the eruption starts with a slow rise, followed by a fast rise phase starting at $t \approx 125$. The rope accelerates to a maximum speed of $V_z \approx 35$ when reaching the upper boundary of the domain at $t = 144$. This is because *external* reconnection takes place in an arch-like current layer which forms at the interface of the expanding ambient field and the preexisting field. This reconnection starts before the flux rope is formed and removes flux from both fields. Thus, the magnetic pressure of the ambient field is gradually diminished and its tension is released.

As the rope rises in the slow phase, *internal* reconnection continues in the current layer below it. In order to see if this reconnection becomes more effective during the evolution of the system, we calculate the maximum value of $J \cdot B/B$, which is proportional to the parallel electric field and indicative of the reconnection rate, in the $y = 0$ plane, within the current layer. The right panel in Figure 2 shows that the reconnection rate grows in time, peaking at $t = 132$, after the onset of the fast rise phase. In order to see why reconnection is increasing, we calculate the maximum relative angle of the field lines which come into contact in the current layer. More precisely, we estimate the full magnetic field vector on both sides of the layer in the plane $y = 0$, and then find the maximum value of the angle between the fields that press against each other. This angle shows gradually increases after $t = 120$. A maximum value of 60 degrees is reached at $t = 132$, coinciding with the peak reconnection rate (Fig. 2). Initially, oppositely directed field lines that come into contact and reconnect make a small relative angle (blue and red field lines in Fig. 1). Later on, this angle increases due to the expansion of the system. The convex, arch-like ends of such field lines are stretched in vertical direction and, thus, the segments of neighbouring field lines that come closer together adopt a larger relative angle and reconnect more efficiently.

As a result of this reconnection, an upward high-velocity outflow is formed below the flux rope. This outflow jet, together with the tension force of the newly reconnected field lines, helps the flux rope to accelerate. Further reconnection (below and above the flux rope) occurs as the accelerated flux rope moves upward and, thus, a runaway situation sets in. External reconnection opens the way for the rope to escape and internal reconnection accelerates it from below. This process is responsible for the transition from the slow to the fast rise phase and for the onset of the full eruption of the rope.

Figure 3 consists of two panels: the left panel shows the topology of field lines around the erupting flux rope. The colormap in the right panel shows the vertical velocity. The isosurface represents the sigmoidal current layer below the rope. Orange field lines have been traced from the flux rope centre, revealing the loop-like shape of the erupting field. Green field lines above the rope belong to the ambient field and have not yet reconnected. They are more azimuthal compared to the cyan field lines which run below the rope. The latter are the result of the internal reconnection. Green and cyan field lines have opposite directions. Blue and red field lines have been traced from a height between the green and cyan field lines, from the side outskirts of the rope. Segments of these field lines are lifting off to be part of the flux rope, while the rest remains at low heights. The rising segments overlay the legs of the bended rope, close to the ends of the current layer. Their downward tension may act against the rise of the rope. Eventually, these field lines reconnect (with the ambient field at large heights and with field lines on the other side of the current layer at low heights), and their tension is released. The white and yellow field lines reconnect around the central area of the current layer. The relative angle between the two sets of field lines close to the reconnection site is now larger than at previous times. Their reconnection yields the cyan field lines, whose tension force sustains the acceleration of the rope, and the lower, arch-shaped green field. Temperature is enhanced at the top of the arch-like structure, due to the collision of the downward reconnection jet with the plasma at the flux pile-up regime below the current layer. A reconnection jet is emitted across the (cyan) reconnected field lines and pushes them upward. Now the flux rope is accelerated with high speed towards the outer atmosphere.

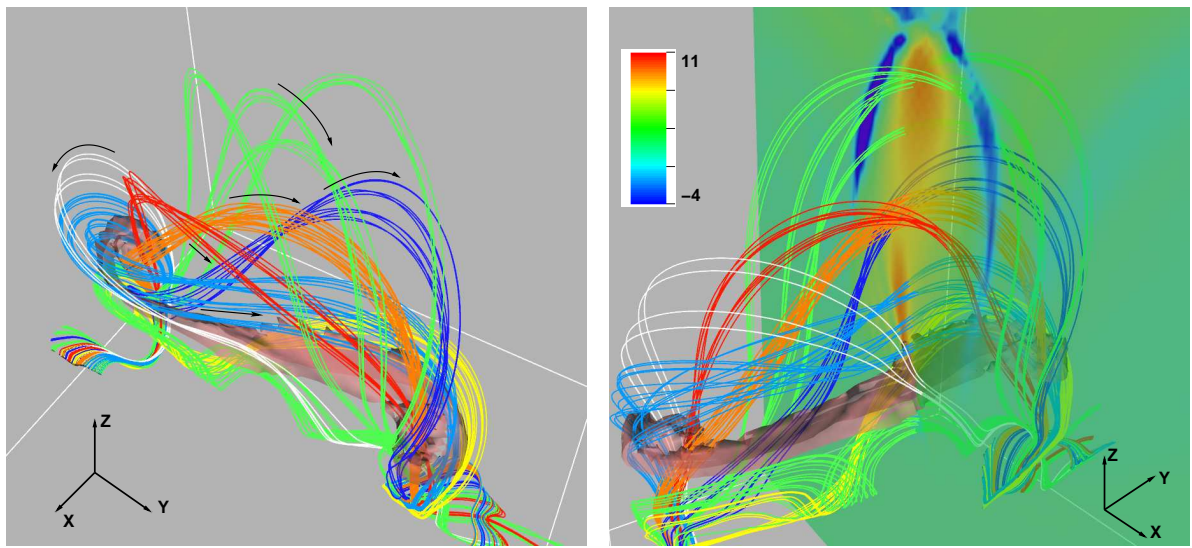

**Fig. 3.** *Left:* Eruption of flux rope (orange lines) and topology of field lines around it at $t = 140$. The arrows show the direction of the field. *Right:* Side view of the same sets of field lines. The colormap shows the vertical velocity at the plane $y = 0$.

## 4. Summary and Discussion

In this Letter, we presented 3D numerical simulations of the emergence of a twisted flux tube from the solar interior into the outer solar atmosphere. The results suggest that formation and eruption of a flux rope, during magnetic flux emergence in active regions, is a generic phenomenon. The formation of the rope is independent of the initial flux and twist of the buoyant magnetic field. Recent simulations (Archontis et al. 2009) have furthermore shown flux rope formation for $\lambda = 20$. However, the characteristics of the evolution of the overall flux system (i.e. ambient field and flux rope) depend on these parameters. Our numerical experiments show that when the emergence occurs into a field-free corona the eruption of the flux rope is confined by the predecessor ambient field. On the other hand, emergence into a magnetized atmosphere results into the full eruption of the flux rope, due to release of the tension of the ambient field.

Further experiments are required to justify the effect of the orientation and the field strength of the pre-existing coronal field in the eruption of the rope.

In the experiments with magnetized atmosphere, the temporal evolution of the height of the rope shows a good agreement with the observed rise profiles of solar eruptions (e.g., filaments, CMEs). Our results suggest that the mechanism responsible for the fast rise and eruption of the rope is the effective reconnection process that occurs underneath the flux rope but also removes the ambient field above it. Additional studies are necessary to investigate the occurrence of instabilities after the flux rope formation and whether they might account as additional drivers of the eruption of the rope. The latter will be particularly important for the experiments where the rope is confined at low heights, with dense plasma trapped in the dips of the rope's field lines. These studies may provide new insights into the formation and eruption of prominences.

In summary, our numerical experiments have made the connection between flux emergence, reconnection and coronal eruptions as a global phenomenon.

*Acknowledgements.* Financial support by the European Comission through the SOLAIRE network (MTRM-CT-2006-035484) is gratefully acknowledged.